\documentclass[12pt,english,dvips]{scrartcl}
\usepackage{color}
\usepackage{babel}
\usepackage[T1]{fontenc}
\usepackage[ansinew]{inputenc}
\usepackage{amssymb}
\usepackage{amsmath}

\begin{document}

\begin{titlepage} \vspace{0.2in} 
\begin{center} {\LARGE \bf 
A Scenario for the Dimensional Compactification \\
in Eleven-Dimensional Space-Time
\\} \vspace*{0.8cm}
{\bf Giovanni Montani
}\\ \vspace*{1cm}

ICRA---International Center for Relativistic Astrophysics\\
Dipartimento di Fisica (G9), 
Universit\`a  di Roma, ``La Sapienza'', 
Piazzale Aldo Moro 5, 00185 Rome, Italy
\vspace*{1.8cm}

 \vspace*{1cm} \\ 

{\bf   Abstract  \\ } \end{center} \indent
We discuss the inhomogeneous multidimensional mixmaster 
model in view of appearing, near the cosmological 
singularity, a scenario for the dimensional compactification
in correspondence to an 11-dimensional space-time.\\ 
Our analysis candidates such a collapsing picture toward
the singularity to describe the actual expanding 3-dimensional
Universe and an associated collapsed 7-dimensional space. 
To this end, a conformal factor
is determined in front of the 4-dimensional metric
to remove the 4-curvature divergences and the resulting Universe
expands with a power-law.inflation.\\
Thus we provide an additional peculiarity of the
eleven space-time dimensions in view of implementing 
a geometrical theory of unification.
\end{titlepage}

The attempt for a geometrical unification of the
fundamental interactions present in Nature leads,
in the classical \cite{C175,C275}, 
as well as supergravity \cite{C78} Kaluza-Klein theories,
to represent the space-time as the direct product
of a generic 4-dimensional manifold
and an internal compact space; such a compact space must  
be homogeneous (to reflect, via its isometries, a
gauge group) and have size of some orders the
Planck length.\\
Though the existence of such an (actually) 
unobservable internal compact space 
can be introduced as an intrinsic feature, due to a 
{\em spontaneous compactification} process 
(i.e. a spontaneous breaking of the Poincar\'e symmetry), 
nevertheless it require a cosmological justification
on the base of a suitable dynamics of dimensional
compactification.

Over the years many efforts were done (see for 
instance \cite{C80,F82}) in order to obtain
a multidimensional cosmology in
which takes place the dynamical decomposition
in terms of an expanding
3-dimensional space phenomenologically
compatible with the Friedmann-Robertson-Walker model) 
and an extra-dimensional one
collapsing to Planckian scales. Here we show how, in
correspondence to an 11-space-time,
the appearance of a
compactification process acquires very general character,
no longer related to the choice of particular models. 

We start by a brief review of the basic
dynamical features characterizing, in vacuum, 
the asymptotic evolution to a spacelike singularity
proper of a generic multidimensional cosmology
(see \cite{BKL70}-\cite{KM95}).

Let us consider a $(d+1)$-dimensional
space-time $(d\ge3)$,
whose associated metric tensor obeys to a dynamics described
by an Einstein-Hilbert action, i.e. 
by the following system of field equations

\begin{equation}
^{(d+1)}R_{ik}=0  ;\qquad  (i,k =0,1,...,d)
\, , 
\label{a}
\end{equation}

where the $(d+1)$-dimensional Ricci tensor
takes its natural form in terms
of the metric components $g_{ik}(x^l)$.\\

In \cite{D86} it is shown that,
within the framework of the Einstein 
theory, the inhomogeneous mixmaster behavior derived in \cite{BKL82},
with respect to generic 3-dimensional cosmologies,
finds a direct generalization in correspondence to any
value of d.\\ 
In a synchronous reference (described by usual coordinates
$(t, x^\gamma)\, \quad \gamma = 1. 2. 3$),
the time evolution of the d-dimensional spatial
metric tensor
$\gamma_{{\alpha}{\beta }}(t, x^\gamma)$ singles out,
near the  cosmological
singularity ($t = 0)$, an iterative structure;
each single stage consists of
intervals of time (Kasner epochs) during which tensor
$\gamma _{\alpha \beta }$ 
takes the generalized Kasner form

\begin{equation}
\gamma_{\alpha \beta}(t, x^\gamma) =
\sum_{i=1}^{d} t^{2p_{l^i}}l^{i}_{\alpha}l^{i}_{\beta}
\, ,
\label{b} 
\end{equation}

where the Kasner index functions
$p_{l^i}(x^{\gamma })$ have to 
satisfy the conditions

\begin{equation}
\sum_{i=1}^{d} p_{l^i}(x^\gamma)\ =
\sum_{i=1}^{d} p_{l^i}(x^\gamma)^2\ = 1 \
\label{c}
\end{equation}

and ${\bf l}^{1}(x^\gamma),...,{\bf l}^{d}(x^\gamma)$
denote $d$ linear independent vectors, whose components
are arbitrary functions
of the spatial coordinates.

In each point of the space, the conditions (\ref{c}) 
define a set of ordered indices
$\{ p_i\}$ ($p_1\le p_2\le ...\le p_d$) which,
from a geometrical point of
view, fix one point in $R^d$, 
lying on a connected portion
of a $(d-2)$-dimensional sphere.
We note that the validity
of the conditions (\ref{c}) 
requires $p_1\le 0, p_{d-1}\ge 0$;
where the equality takes place only for the values
$p_1 = . . .= p_{d-1} = 0, p_d =1 $.

As shown in \cite{DHS85,HJS87} (see also \cite{K93,KM95},
each single step of this 
iterative solution results to be stable, in a given 
point of the space, if
take place there the following conditions:

\begin{equation}
\forall (x^1,...,x^d) :\qquad
\alpha_{ijk}(x^\gamma) > 0 \qquad (i\neq j, i\neq k, j\neq k)\qquad
(i,j,k=1,...,d) \, , 
\, , 
\label{d}
\end{equation} 

where the quantities $\alpha _{ijk}(x^\gamma)$
are defined, in each
space point, by expressions of the form:

\begin{equation}
\alpha_{ijk} = 2p_i + \sum_{l\ne i,j,k} p_l\;\qquad
(i\ne j, i\ne k, j\ne k),\qquad
(i,j,k=1, ...,d)
\label{e}
\end{equation}

and take values in the available domain for
the ordered indices $\{ p_i\}$. 

It can be shown \cite{DHS85,HJS87}  
that ,for $3\le d\le 9$ 
at least the smallest of the quantities 
(\ref{e}), i.e. $\alpha_{1,d-1,d}$ results to be
always negative
(excluding isolated points $\{p_i\}$ in which it vanishes); 
while for $d\ge 10$  there exists an open region of the
$(d-2)$-dimensional Kasner sphere where this same
quantity takes positive values, 
the so-called Kasner Stability Region (KSR).\\
As a consequence, for $3\le d \le 9$, the
evolution of the
system to the singularity consists of an infinite
number of Kasner epochs; instead for $d\ge 10$,
the existence of the KSR, 
implies a profound modification in the asymptotic dynamics. 
In fact the (reliable) indications presented
in \cite{D86,KM95} in favor of the ``attractivity 
of the KSR, imply that in each space point
(excluding sets of zero measure)
a final stable Kasner-like regime appears.

Finally we stress that, in correspondence
to any value of $d$, 
the considered iterative scheme contains the right number of
$(d+1)(d-2)$ physically arbitrary functions of the
spatial coordinates,
required to specify generic initial conditions
(on a non-singular
spacelike hypersurface); therefore this piecewise solution 
describes the asymptotic evolution
of a generic inhomogeneous multidimensional
cosmological model.

Now we show how, for $d = 10$ the Kasner stability
region is characterized by a peculiar feature which
has relevant dynamical implications
for a dimensional reduction scenario  
toward the singularity.

First we rewrite,
in terms of the ordered Kasner indices $\{p_i\}$, 
the conditions which
define in $R^d$ the $(d-2)$-dimensional allowed domain, i.e. 

\begin{equation}
\sum_{i=1}^{d} p_i \ = \sum_{i=1}^{d} p_i^2 \ =1 ;\qquad
p_1\le p_2\le ...\le p_d \, . 
\label{f}   
\end{equation}

Within such a domain, the KSR
is defined by the following conditions: 

\begin{equation}
\alpha_{1,d-1,d} = 2p_1 + p > 0; \qquad
p = \sum_{i=2}^{d-2} p_i
\, ; 
\label{g}
\end{equation}

in fact, the validity of this inequality
ensures the positiveness of all the quantities 
$\alpha_{ijk}, (i\neq j,i\neq k, j\neq k),
(i,j,k=1,...,d)$.\\ 
We observe that those sets $\{p_i\}$ which verify
the condition $\alpha_{1,d-1,d} = 2p_1 + p = 0$,
and therefore (for $d\ge 10$) lay at the
boundary of the Kasner stability
region, are fixed points with respect to the
iteration of the multidimensional map (associated with replacing of Kasner
indices). 
It is worth noting that constitutes an exception
the point $p_1 = ...= p_{d-1} =0$,$p_d = 1$ which,
although is a fixed one for any value of
$d$, nevertheless it does not lay (for $d\ge 10$) at the
boundary of the Kasner stability
region, remaining in this sense an isolated fixed point. 

With clear reference to our leading idea,
let us now search for points
$\{p_i\}$ in the allowed domain having the
following structure:

\begin{equation}
 p_1 = p_2 = p_3 = -X ; \qquad
p_4 = p_5 = ... = p_d = Y ; \qquad (X\ge 0, Y > 0);\qquad (d > 3) \
\, . 
\label{h}
\end{equation}

By the conditions (\ref{f})-(\ref{g}) we obtain the following
simple algebraic system in the variables $X, Y$:

\begin{equation}
-3X + (d-3)Y = 1 ; \qquad 
3X^2 + (d-3)Y^2 = 1 ; \quad
(X\ge 0, Y > 0); \qquad (d > 3)
\, ,
\label{i}
\end{equation}

whose solution gives us the explicit
expressions of $X$ and $Y$ as
functions of $d$;
so we define, in the allowed domain, a point
$\{p_i(d)\}$ having the form
\footnote{We note that the existence
of such points in the case $d=9$ and $d=10$
was first pointed out
in \cite{DHS85}
with reference to a different purpose.}:
\begin{subequations}
\begin{align}
& p_1 = p_2 = p_3 = \frac{1}{d} \
[1-\sqrt{\frac{1}{3} (d-1)(d-3)}];  \\ 
& p_4 = p_5 = ... = p_d = \frac{1}{d(d-3)} \ 
[d-3 + \sqrt{3(d-1)(d-3)}]
\, . 
\end{align}
\label{l}  
\end{subequations}
In correspondence to these ``special'' points
the quantity
$\alpha_{1,d-1,d}$ takes the following explicit
expression
as function of $d$:
\begin{equation}
\alpha_{1,d-1,d}[p_i(d)] \equiv  \alpha^*_{1,d-1,d}(d) = \
\frac{1}{d} \
[d-1 - \frac{d+3}{d-3 } \
\sqrt{\frac{1}{3} (d-1)(d-3)}]
\, . 
\label{m}
\end{equation}
It can be verified the validity of the following
statements:\\ 
a)$\; For \ \ d = 4, \alpha^*_{1,3,4} = 0 \ \
in \ \ (p_1 = p_2 =
p_3 = 0, p_4 = 1)$ \\ 
b)$\; For \ \ 5\le d\le 8, \alpha^*_{1,d-1,d}< 0 $ \\ 
c$\; For \ \ d= 9, \alpha^*_{1,8,9}= 0 \ \ in \ \ (p_1 = p_2 = p_3 = -1/3;
 p_4 = p_5 = ...= p_9= 1/3)$ \\ 
d)$\; For \ \ d\ge 10, \alpha^*_{1,d-1,d} > 0$

Thus we find the key result that, 
for $d\ge 10$, the
points of the form (\ref{l}) 
belong always to the KSR.
This fact acquires particular relevance
for the dimensional reduction 
since it is clear that,
due to the continuity, in a small enough neighborhood
of such points
(interely within the KSR)
the Kasner indices $\{p_i\}$ have to 
conserve the same structure
with three negative values and all the other positive ones
(although, in general, we have no longer any equality
condition in each of these two groups of values).

Let us now show that for $d = 10$
there exists a whole connected domain in the 
KSR which possesses such a structure, i.e. 
is constituted by points $\{p_i\}$ for which always
three indices are negative and the remaining seven
all positive ones.\\ 
For $d = 10$, the point
of the form (\ref{l}) reads explicitly

\begin{equation}
 p_1 = p_2 = P_3 = (1-\sqrt{21})/10; \qquad
 p_4 = p_5= ...= p_{10}= (7+3\sqrt{21})/70 \
\label{n}
\end{equation}

and the quantity $\alpha^*_{1,9,10}$
takes the positive value
$(63-13\sqrt{21})/70$.\\ 
In a small neighborhood
of the point (\ref{n}), the KSR 
have to contain points with three negative indices.
We observe that, as far as we deal with a connected region,
the presence of points
$\{p_i\}$ having two or four
(in case till to $d-2$) negative indices, 
implies that there exist curves (interely inside such a region),
which joint points of such a kind to the one (\ref{n}).\\ 
However, any of such curves has to 
necessarily cross the hyperplanes
$p_3 = 0$ or $p_4 = 0$ in correspondence to points
of the form

1) $[(p_1 < 0);\ \ (p_2\le 0);\ \ (p_3 = 0);\ \ (p_4,...,p_8\ge 0);\ \
   (p_9, p_{10} > 0)]$

2) $[(p_1 < 0);\ \ (p_2, p_3\le 0);\ \ (p_4 = 0);\ \ (p_5,...,p_8\ge 0);\ \
 (P_9, P_{10} > 0)]$
 
where all the above non-vanishing values 
have to be regarded as unspecified generic ones.
Of course sets of values of the form 1) and 2)
define points belonging to the KsR,
only if the conditions (\ref{f}) and (\ref{g})  
take place simultaneously. 
But (at least) one index, $p_3$ or $p_4$,
must be zero and therefore such conditions reduce
exactly to those ones relative
to a 9-dimensional space.
Since, as discussed before, in correspondence to 
$d = 9$ there exist no points in the allowed domain
satisfying the condition (\ref{g}) and then 
we can conclude that for $d = 10$
it exists a whole connected portion of 
the KSR which does not contain
points of the form 1) or 2).\\ 
Thus, in agreement to the previous reasoning,
it follows that, for $d = 10$,
such a connected region is really  constituted
only by sets 
$\{p_i\}$ having three indices always negative and all the 
remaining seven positive ones, i.e of the form:

3) $[(p_1, p_2, p_3 < 0); \ \ (p_4,...,p_{10} > 0)]$

It is easy to recognize that such a
connected region should have a boundary
constituted of fixed points.
In analogy to the above proof, we can 
show that, for $d = 10$,
also these fixed points possess the
same structure 3), with the only exception of the special
point

\begin{equation}
  p_1 = p_2 = p_3 = -1/3;\qquad p_4 = 0;\qquad 
  p_5 = p_6 = ...= p_{10} = 1/3.
  \, . 
\label{p}   
\end{equation}

However, we stress that, to the 10-dimensional 
KSR belongs the point 

\begin{equation} 
p_1 =...= p_4= -(\sqrt{27/2} - 1)/10 
\, \quad 
p_5 =...= p_10 = (1 + \sqrt{6})/10
\, ,
\label{neg}
\end{equation}

which, in fact, corresponds to 

\begin{equation} 
{\alpha}_{1,9,10} =
(9 - 7/2\sqrt{6})/10 > 0
\, .
\label{al}
\end{equation} 

Here, the relevant feature relies on that,
the region having the property above outlined is an isolated one
and the space regions where three and four
indices are negative 
(i. e. the corresponding number of dimensions expand), 
are separated by the 2-dimensional surface defined via
the conditions

\begin{equation}
\{ \; p_1(x^{\gamma }) = p_2(x^{\gamma }) =
p_3(x^{\gamma }) = -1/3 \, \quad 
p_4(x^{\gamma }) = 0 \, \quad 
p_5(x^{\gamma }) = .... = 
p_8(x^{\gamma }) = 1/3
\, ; 
\end{equation}

indeed the remaining indices
$p_9$ and $p_{10}$ are obliged to the value
$1/3$ by the conditions (\ref{f}) 
\footnote{ The picture is made a bit more
complicated by the existence of
points of space where never appear a stable set of indices.}. 

We conclude this analysis by observing
that the above properties of
the KSR, as  derived for
$d = 10$, do not hold for higher dimensional cases.\\ 
In fact, for $d = 11$, the KSR 
still never meets the
hyperplane $p_3 = 0$, but now it contains
points of the hyperplane
$p_4 = 0$ in correspondence to a small enough
neighborhood of the point 

\begin{equation} 
  p_1 = p_2 = p_3= (1-\sqrt{21})/10;\qquad
  p_4 = 0,\qquad 
  p_5 = p_6 = ...= p_{11} = (7+3\sqrt{21})/70
 \, , 
\label{q}  
\end{equation} 

for which $\alpha_{1'10'11}$
coincides with $\alpha^*_{1,9,10}( > 0)$.\\ 
Thus, the iteration of our discussion, 
for increasing values of $d$, allows
to say that the KSR can never contain
points $\{p_i\}$ with only two negative indices; 
but a whole connected portion of the KSR,
having three negative Kasner indices,
no longer exists.
Thus for $d > 11$ the portions of the space where 
three indices are negative  correspond to open sets which 
lay directly in contact with those ones where 
four indices become negative, via the
$d - 1$-dimensional hypersurface
$p_4(x^{\gamma }) = 0$.

Now we describe the dynamical implications of
the results above obtained about the KSR, 
with respect to the asymptotic evolution
toward the singularity.\\
We see how, for $d = 10$,
due to the structural feature of the KSR,
during the asymptotic evolution, 
in each space point (of the spatial domains,
bounded by 2-dimensional surfaces), 
a stable Kasner-like regime take place. 
The dynamics is characterized by a natural decomposition
in terms of three expanding directions
(associated in that point to the three
negative Kasner indices)
and the collapse of the remaining seven ones
(associated to 
the positive indices).\\ 

However it is important to observe that,
as a consequence of the ``oscillatory
behavior'' (induced, on the spatial dependence
of the Kasner indices, by the
map iteration, \cite{M95}-\cite{KM95}), 
the three expanding directions, as 
well as the collapsing ones, change through space.

Though in this asymptotic scheme we
obtain a 3-dimensional expanding Universe, 
nevertheless the associated
4-dimensional space-time curvature
diverges as $1/t^2$
($t\rightarrow 0$).\\ 
This aspect would make unsuitable the implementation of
any actual cosmology in this dynamical context. 
However this divergence
can be removed by, first, considering the following
time coordinate transformation

\begin{equation}
T = \frac{1}{t^{1/X}} \, \quad X = const. \, \quad X > 0 
\label{r} 
\end{equation}

and then taking the conformal factor
in front of the 4-dimensional metric which restores 
a synchronous reference; hence the line element of our model
rewrites 

\begin{equation}
ds^2 = \frac{X^2}{T^{2 + 2X}}\left( dT^2 - \sum_{i=1}^{3} 
\frac{1}{X^2}T^{2 + 2X(1 - p_{l^i})}l^i_{\alpha }
l^i_{\beta} dx^{\alpha }dx^{\beta }\right)
- \sum_{j=4}^{10}T^{-2Xp_{l^j}}l^j_{\alpha } l^j_{\beta }dx^{\alpha}dx^{\beta}  
\, . 
\label{s}
\end{equation}

Of course,this expression of the metric
must be referred to those regions of
the space where the indices
$p_{l^1}$, $p_{l^2}$ and
$p_{l^3}$ assume the three allowed negative values.\\
As $T\rightarrow \infty$
the 4-dimensional curvature acquires a vanishing 
behavior.and 
we get a power-law inflation characterizing
the 3-dimensional Universe expansion. This latter feature is 
particularly interesting because a power-law expansion stretches
the inhomogeneities out of the (redefined) physical 3-horizon. 
It is worth noting that to get
a real dimensional reduction,
from extra-dimensions collapse, 
impliess certain topology conditions on the model or mechanisms
which prevent the direct observation.
We also infer that, in our generic picture,
the isometries, underlying the gauge fields representation,
have to be recognized locally.

We conclude our analysis by
stressing how the idea proposed in
this paper, represents an additional
argument confirming the privileged
character of the 11-dimensional space-time in view
of the implementation of a geometrical unification theory. 
 
In fact, as discussed in \cite{W81}, the
11-dimensional space-time turns
out to be the most natural choice
when considering the construction of a
realistic Kaluza-Klein theory.
This claim is supported, in first place,
by a phenomenological argument, based on the necessity
of representing (in the
zero mode approximation) 
the right massless gauge fields responsible for
the fundamental particles interactions. 
Since in the Standard Model the
bosonic component is described by the gauge
group $SU(3)\otimes SU(2)\otimes U(1)$,
then the symmetry group of the extra-compact
internal space must contain it.
Now it is possible to show that
the minimum dimension of a manifold, having the
$SU(3)\otimes SU(2)\otimes U(1)$ symmetry, 
is a 7-dimensional one, i.e. 
$CP^2\times S^2\times S^1$. 
We see how this phenomenological constraint indicates
the 11-dimensional space-time
as the lowest-dimensional admissible one for a
realistic Kaluza-Klein theory.\\ 
Furthermore, it represents a remarkable fact
that eleven is even the
maximum number of dimensions for which
supergravity theory is completely consistent. 
In fact,for more than eleven
dimensions, the gravitino, which is a Rarita-Schwinger
spin-vector, would have more than $128$ degrees of freedom and,
once reduced in four dimensions, it would lead to
a supergravity theory
containing massless particles with spin greater than two. 
But there are strong arguments in favor of the idea that
the coupling
of such kind of particles with gravity leads to inconsistency.

On the base of all these considerations
it looks really remarkable that, in
correspondence to
an 11-dimensional space-time,
this large
number of peculiar aspects
takes place in 
favor of the implementation of a
geometrical Unification Theory; 
it leads us to argue that they these features
might constitute more than mere coincidences. 

\vspace{1cm}

V. A. Belinski is thanked for having attracted our attention
on the point of view here addressed.\\ 
A. A. Kirillov and R. Ruffini are thanked for the
interesting discussions on this subject and their
valuable advice.

\end{document}